\documentclass[sigconf]{acmart}
\acmConference[ICSE'24-Poster]{International Conference on Software Engineering --- Invited Posters}{Apr.'24}{Lisbon, Portugal}

\newif\ifDEBUG
\DEBUGtrue

\AtBeginDocument{%
  \providecommand\BibTeX{{%
    \normalfont B\kern-0.5em{\scshape i\kern-0.25em b}\kern-0.8em\TeX}}}

\usepackage{threeparttable}
\usepackage{tabularx}
\usepackage{tabulary}
\usepackage{colortbl}
\usepackage{multicol}
\usepackage{multirow}
\usepackage{xurl}  
\usepackage{hyperref}

\usepackage{pgf-pie}  
\usepackage{enumitem}
\usepackage{xspace}

\usepackage{soul}

\ifDEBUG
    \newcommand{\JD}[1]{\textcolor{purple}{[JD says:#1]}}

\else
    \newcommand{\JD}[1]{}

\fi

\usepackage{tcolorbox}

\makeatletter
\def\cl@chapter{}
\makeatother
\usepackage{cleveref}

\crefformat{section}{\S#2#1#3}
\crefname{figure}{Figure}{Figures}
\crefname{appendix}{Appendix}{Appendices}
\crefname{table}{Table}{Tables}
\crefname{algorithm}{Algorithm}{Algorithms}
\crefname{listing}{Listing}{Listings}
\crefname{theorem}{Theorem}{Theorems}
\crefname{thm}{Theorem}{Theorems}
\crefname{lemma}{Lemma}{Lemmata}
\crefname{equation}{Eqt.}{Eqts.}
\crefformat{Grammar}{Grammar #1}

\newcommand{\ie}{\textit{i.e.,} }
\newcommand{\eg}{\textit{e.g.,} }
\newcommand{\etal}{\textit{et al.}\xspace}

\newcommand{\myparagraph}[1]{\vspace{0.05cm}\textbf{#1}}
\newcommand{\mysubsection}[1]{\vspace{0.07cm}\noindent\textbf{#1:}}

\newcommand{\code}[1]{{\small\texttt{#1}}\xspace}

\usepackage[font=small,skip=0pt]{caption}
\usepackage{xspace}

\newcommand{\MyTitle}[1]{}
\renewcommand{\MyTitle}{Understanding the Impact of Data Privacy Regulations on open-source Software}
\renewcommand{\MyTitle}{Legal compliance Meets Open-Source: A First Look in the Context of GDPR}
\renewcommand{\MyTitle}{An Exploratory Study of GDPR Compliance in Open-Source Projects}
\renewcommand{\MyTitle}{Perceptions of GDPR in Open-Source Software}
\renewcommand{\MyTitle}{Open-Source Perceptions of Privacy: A Case Study in GDPR Compliance}
\renewcommand{\MyTitle}{Perceptions of GDPR Compliance in Open-Source Software Development}
\renewcommand{\MyTitle}{Understanding the Impact of the General Data Protection Regulation (GDPR) on Open-Source Software Development}
\renewcommand{\MyTitle}{The General Data Protection Regulation (GDPR) in Open-Source Software}
\renewcommand{\MyTitle}{A First Look at the General Data Protection Regulation (GDPR) \\ in Open-Source Software}

\title{\MyTitle}

\author{Lucas Franke}
    \affiliation{%
      \institution{Virginia Tech}
      \country{Blacksburg, VA, USA}}
    \email{lfranke@vt.edu}
    
\author{Huayu Liang}
    \orcid{0000-0003-4193-3985}
    \affiliation{%
      \institution{Virginia Tech}
      \country{Blacksburg, VA, USA}}
    \email{huayu98@vt.edu}
    
    \author{Aaron Brantly}
    \orcid{0000-0003-4193-3985}
    \affiliation{%
      \institution{Virginia Tech}
      \country{Blacksburg, VA, USA}}
    \email{abrantly@vt.edu}
    
    \author{James C. Davis}
    \orcid{0000-0003-2495-686X}
    \affiliation{%
      \institution{Purdue University}
      \country{W Lafayette, IN, USA}}
    \email{davisjam@purdue.edu}
    
    \author{Chris Brown}
    \orcid{}
    \affiliation{%
      \institution{Virginia Tech}
      \country{Blacksburg, VA, USA}}
    \email{dcbrown@vt.edu}

\begin{document}

\begin{abstract}

This poster describes work on the General Data Protection Regulation (GDPR) in open-source software.
Although open-source software is commonly integrated into regulated software, and thus must be engineered or adapted for compliance, we do not know how such laws impact open-source software development.

We surveyed open-source developers (N=47) to understand their experiences and perceptions of GDPR.
We learned many engineering challenges, primarily regarding the management of users' data and assessments of compliance.
We call for improved policy-related resources, especially tools to support data privacy regulation implementation and compliance in open-source software.

\end{abstract}

\maketitle

\vspace{-0.2cm}
\section{Problem Statement} \label{sec:Introduction}
\vspace{-0.1cm}

Software products collect user data to enhance user experiences through personalized, machine learning-enabled~\cite{pew2019} application behaviors~\cite{bucklin2009click}. 
This may both benefit users and threaten their well-being, \eg swaying elections in the USA~\cite{nymag2019}. 
%
%
To protect their citizens, over 100 governments worldwide are developing \emph{data privacy laws and regulations}
to constrain how citizens' personal data is collected, processed, stored, and saved~\cite{unctad}.
A landmark example is the EU's General Data Protection Regulation (GDPR), which grants rights to citizens for the handling of their data~\cite{gdpr2018}.

Data privacy regulations create challenging software requirements because they entail both technical and legal expertise. 
Software developers may have limited legal knowledge~\cite{verdon2006security,moquin2016roles} and receive minimal training~\cite{allan2007reskilling,holst2017liability}.
However, there has been limited study of how such laws affect the software development process.
The few existing studies have been of commercial software development~\cite{alhazmi_arachchilage_2021,doi:10.1080/01972243.2019.1583296}; we lack knowledge of the effects of GDPR and other regulations on open-source software (OSS) development.

You may be surprised that regulatory compliance is a factor in OSS development, as most OSS licenses disclaim legal responsibility. 
However, users and developers of OSS may still desire regulatory compliance.
We note three examples.
(1) A majority of OSS is developed for commercial use~\cite{octoverse2022} and may require standards or regulatory compliance. 
(2) Companies who integrate OSS components in their software supply chains~\cite{okafor2022sok} may request the addition of compliance requirements, and the developers may service these requests.
(3) Companies may find the (free) price of OSS compelling and undertake their own compliance analysis~\cite{STOKES2012481}.
Standards such as IEC 61508--Part 3 include provisions for doing so~\cite{IEC61508-3}. 
In a 2023 survey of $\sim$1700 codebases across 17 industries, Synopsys found OSS in 96\% of the codebases and reported an average contribution of 75\% of the code in each codebase~\cite{synopsys2023report}.

\vspace{0.05cm}
\ul{\textit{We therefore explore GDPR compliance in open-source software.}}

\vspace{-0.2cm}
\section{Approach}
\vspace{-0.1cm}

GDPR compliance is an unstudied topic in OSS.
We therefore adopt an exploratory methodology, focused on qualitative data, to provide a characterization and identify phenomena of interest.
%
%
We followed a four-step approach aligned with the framework analysis methodology~\cite{frameworkanalysis}. 
Since we studied human subjects, our Institutional Review Board (IRB) provided oversight.


{
\begin{table*}[th!]
\small
    \centering
    \caption{
    Sample questions from pilot study and survey.
    The final column notes the inter-rater agreement score for these themes (Cohen's $\kappa$). 
    }
    \begin{tabular}{cc|ccc}
        \toprule
        \textbf{Interview Question} & \textbf{Example Codes} & \textbf{Survey Question} & \textbf{Example Codes} & \textbf{$\kappa$} \\
        \toprule
        \multirow{4}{0.28\linewidth}{What impact, if any, do you believe the GDPR has had on data security and privacy?} & \multirow{4}{0.12\linewidth}{data privacy, rights to users, data collection} & \multirow{4}{0.28\linewidth}{What impact, if any, do you believe the GDPR and similar data privacy regulations have had on data security and privacy?} & \multirow{4}{0.2\linewidth}{\small data privacy, data processing, data collection, insufficient information, data breach, fines} &  \\
        & & & & \\
        & & & & \\
        & & & & 0.736 \\ \midrule
        \multirow{3}{0.280\linewidth}{What GDPR concepts do you find the most difficult or frustrating to implement?} & \multirow{3}{0.12\linewidth}{None, data minimization, embedded content} & \multirow{3}{0.28\linewidth}{What GDPR concepts do you find the most difficult or frustrating to implement?} & \multirow{3}{0.2\linewidth}{\small privacy by design, data minimization, cost, data processing, user experience}
        &  \\
        & & & & \\
        & & & & \\ \midrule
        \multirow{3}{0.280\linewidth}{Have you sought legal consultation on GDPR-related issues, and if so, how did that affect your development process?} & \multirow{3}{0.12\linewidth}{Yes/No; no effect, negative effect (time)} & \multirow{3}{0.280\linewidth}{Have you sought legal consultation on GDPR-related issues, and if so, how did that affect your development process?} & \multirow{3}{0.2\linewidth}
        {\small Yes/No; no effect; positive; negative (cost, time, data storage, data processing,...)}
        & \\
         & & & & \\
        & & & & 0.514 \\
        \bottomrule
    \end{tabular}
    \label{tab:SurveyInstrument}
\end{table*}
}

\mysubsection{Step 1: Pilot Study and Data Familiarization}
To formulate an initial thematic framework for our qualitative analysis, we conducted semi-structured pilot interviews with three developers.

Two researchers coded the interview transcripts to extract themes (examples in~\cref{tab:SurveyInstrument}).
Participants highlighted the challenges with implementing GDPR requirements in open-source software, including (1) legal consultation, and (2) compliance assessment.

\mysubsection{Step 2: Survey Design}
Our pilot study findings informed our survey design. 
As summarized in~\cref{tab:SurveyInstrument}, we asked short-answer questions about GDPR in OSS.
We asked respondents about the perceived impact of the GDPR on data privacy, difficult concepts to implement, and GDPR compliance assessment. 

\mysubsection{Step 3: Participant Recruitment}
We wanted respondents with experience implementing GDPR in OSS.
We searched GitHub for repositories with GDPR-related pull requests (searching for ``GDPR'' typically yields English-language PRs related to GDPR compliance). 
We emailed developers who authored or commented on these PRs.
From 98 emails we received 5 responses (5\% response rate).
We then made broader calls for participation on Twitter and Reddit.

We received a total of 47 responses that self-reported experience implementing GDPR compliance.
Participants have a median of $\sim$5 years of OSS development experience and 6 years of industry experience.
Participants contributed to OSS projects such as Ansible,
Django, Kubernetes, PostGreSQL,
and Microsoft Cognitive Toolkit.


\mysubsection{Step 4: Data Analysis}
%
To analyze our survey results, we used an open coding approach.
Two analysts independently performed a manual inspection of responses, categorizing responses based on the themes derived from our pilot study.
If a new theme arose, the analysts discussed before adding it.
We used Cohen's kappa ($\kappa$) to measure inter-rater agreement (\cref{tab:SurveyInstrument}).


\section{Results so Far}
\vspace{-0.1cm}

\mysubsection{Positive Views}
Six participants had positive perceptions of the GDPR. 
For example, they said 
  ``\textit{the risk of incurring and paying out hefty fines has made companies take privacy and security more proactively}'' (P30),
  that GDPR brings ``\textit{awareness to the importance about privacy}'' (P45),
  and
  that
  ``\textit{data integrity is ensured}'' (P47).
  These responses reflect the intentions of the GDPR. 
  
\mysubsection{Negative Views}
16 participants reported negative views.
These responses primarily focused on three issues: cost, organizations, and enforcement.
For costs, respondents noted that  implementing GDPR requirements is expensive and burdensome.
Participants said that compliance is ``\textit{costly for many companies}'' (P16) is ``\textit{too expensive}'' (P24), and ``\textit{the cost of protection should not go over the cost of consequence of data breach...GDPR [isn't] worth the time}'' (P46).
For organizations, participants reported a negative impact of the GDPR on companies and organizations.
They mentioned that GDPR compliance ``\textit{weakens small and medium-sized enterprises}'' (P15), ``\textit{threatens innovation}'' (P18), and ``\textit{fails to meaningfully integrate the role of privacy-enhancing innovation and consumer education in data protection}'' (P23).
P46 added that the GDPR is ``\textit{a lot of headache...jobs for lawyers at the expense of people who are trying to solve real problems}''.
For enforcement, one subject said ``\textit{there is a large gap in GDPR enforcement among member states} (P17) and another had observed changes over time --- ``\textit{the trend...is an increase in the number of times and the amount of fines}'' (P18).

\mysubsection{Engineering Challenges}
We summarize three categories here.

\textit{(1) Software Design:}
P21 felt that GDPR compliance reduced the quality of their application's design: ``\textit{the principle of minimum scope was not observed}''.
One participant said GDPR work showed ``\textit{things we had not considered before}'', \eg ``\textit{logging functionality}'' and ``\textit{access restrictions}''.
P17 observed difficulties with new technologies: ``\textit{GDPR's requirements are essentially incompatible with big data, artificial intelligence, blockchain, and machine learning}''.

\textit{(2) Legal Compliance:}
11 respondents reported consulting with legal teams for GDPR compliance.
Most of them (7) lamented this need, stating ``\textit{it slows things down as code has to be reviewed and objectives revised}'' and ``\textit{it impacted our approach to the SDLC}'' (P1), ``\textit{it's a bit of a headache}'' (P24), and the development process was affected ``\textit{if the development is production software}'' (P27) already deployed to users.
P47 stated ``\textit{open-source projects can't afford even to sustain maintainers, not even speaking about legal team}''. 

\textit{(3) Validation}
Lacking legal counsel, many respondents felt responsible for evaluating ``\textit{legality}'' (P18) and ``\textit{integrity and confidentiality}'' (P23) of user data processing and storage.
P24 responded developers have to ``\textit{consider whether you really need all the data you collect}'' and P38 advised to ``\textit{get your consent in order}''. 
P18 added there is ``\textit{really no good way}'' to evaluate compliance.

\bibliographystyle{ACM-Reference-Format}
\bibliography{davis,main}

\end{document}